\documentclass[app]{revtex4-1}

\usepackage{amssymb}
\usepackage{amsmath,bm}

\usepackage[pdftex]{graphicx,color}
\usepackage[english]{babel}
\usepackage{hyperref}
\usepackage{epstopdf}

\begin{document}

\title{Emission of intense resonant radiation by dispersion-managed Kerr cavity solitons}

\author{Alexander U. Nielsen$^{1,2}$}
\email{anie911@aucklanduni.ac.nz}
\author{Bruno Garbin$^{1,2}$}
\author{St\'ephane Coen$^{1,2}$}
\author{Stuart G. Murdoch$^{1,2}$}
\author{Miro Erkintalo$^{1,2}$}
\email{m.erkintalo@auckland.ac.nz}

\affiliation{$^1$The Dodd-Walls Centre for Photonic and Quantum Technologies, New Zealand}
\affiliation{$^2$Department of Physics, The University of Auckland, Auckland 1010, New Zealand
}

\begin{abstract}
	We report on an experimental and numerical study of temporal Kerr cavity soliton dynamics in dispersion-managed fiber ring resonators. We find that dispersion management can significantly magnify the Kelly-like resonant radiation sidebands emitted by the solitons. Because of the underlying phase-matching conditions, the sideband amplitudes tend to increase with increasing pump-cavity detuning, ultimately limiting the range of detunings over which the solitons can exist. Our experimental findings show excellent agreement with numerical simulations.
\end{abstract}
\maketitle

\section{Introduction}
\noindent Coherently-driven, dispersive Kerr resonators have attracted considerable attention over the last few years~\cite{kippenberg_microresonator-based_2011, coen_temporal_2016, barland_temporal_2017, pasquazi_micro-combs:_2017, kippenberg_dissipative_2018}. Interest has in particular ensued from the realization that, despite being entirely passive and void of any saturable absorbers, such devices can sustain ultrashort pulses of light known as temporal Kerr cavity solitons (CSs)~\cite{wabnitz_suppression_1993}. These CSs arise through a double-balance: coherent driving compensates for the resonator losses, while (anomalous) group-velocity dispersion (GVD) is balanced by the Kerr nonlinearity. They were first observed in a macroscopic fiber ring resonator in 2010~\cite{leo_temporal_2010}, and proposed as bits for all-optical storage and processing applications~\cite{jang_temporal_2015, jang_all-optical_2016}. In 2014, they were also observed in a monolithic microresonator~\cite{herr_temporal_2014}, and are today understood to underpin the generation of coherent ``Kerr'' optical frequency combs in such devices~\cite{coen_modeling_2013, chembo_spatiotemporal_2013, parra-rivas_dynamics_2014,yi_soliton_2015, joshi_thermally_2016, webb_experimental_2016}. This latter observation has especially made CSs relevant to a wide range of contemporary photonics applications~\cite{suh_microresonator_2016, marin-palomo_microresonator-based_2017, suh_soliton_2018, trocha_ultrafast_2018, lucas_spatial_2018}, stimulating numerous works aimed at better understanding the solitons' dynamics and characteristics~\cite{anderson_observations_2016, brasch_photonic_2016, anderson_coexistence_2017, yi_single-mode_2017, brasch_self-referenced_2017, pfeiffer_octave-spanning_2017, wang_universal_2017, cole_soliton_2017, obrzud_temporal_2017, wang_stimulated_2018, hendry_spontaneous_2018}.

While temporal CSs have so far been predominantly studied in longitudinally homogeneous resonators made out of a single uniform waveguide, a handful of studies have also considered longitudinally non-uniform resonators that exhibit parameter variations over a single round trip. Fibre ring resonators constructed of two different fibers have notably been used to experimentally study CS dynamics in the regime of small net (overall) \emph{anomalous} GVD, where effects of higher-order dispersion are more pronounced~\cite{wang_universal_2017, jang_observation_2014}. The impact of such dispersion management on the CS characteristics has also been systematically studied using numerical simulations~\cite{bao_stretched_2015}. These simulations show that, in analogy with mode-locked lasers~\cite{tamura_77-fs_1993}, dispersion management represents a promising avenue for shortening the temporal width of the CSs that can be generated.

In parallel with investigations related to CSs, several studies have considered the nonlinear dynamics in longitudinally non-uniform resonators featuring a net \emph{normal} GVD~\cite{copie_competing_2016, conforti_parametric_2016, copie_dynamics_2017, copie_modulation_2017, huang_quasi-phase-matched_2017, garbin_experimental_2017}. These studies have been predominantly motivated by the fact that the dispersion variations can magnify (and extend) the instabilities that arise due to the intrinsic cavity periodicity, permitting e.g. controlled investigations of the competition between Turing and Faraday pattern forming mechanisms~\cite{copie_competing_2016, conforti_parametric_2016, copie_dynamics_2017}. But CSs --- which only manifest themselves in the region of \emph{anomalous} GVD --- are of course also known to be affected by the cavity periodicity through the emission of Kelly-like resonant radiation sidebands~\cite{jang_observation_2014, wang_universal_2017}, and so a natural question that arises is whether (and how) dispersion management influences this process. In addition to being of fundamental interest, the fact that dispersion management could allow for the generation of ultrashort CSs in fiber resonators and low-Q microresonators~\cite{bao_stretched_2015} renders the question important also from an applied perspective. Yet, to the best of our knowledge, no detailed experimental or theoretical work has hitherto been reported.

Here we report on an experimental and numerical study of temporal Kerr CS dynamics in dispersion-managed fiber ring resonators. We find that, as for parametric instabilities in resonators with normal net GVD~\cite{conforti_parametric_2016}, dispersion management can considerably magnify the Kelly-like resonant radiation sidebands emitted by CSs. Moreover, we show that, due to the underlying phase-matching conditions, the amplitude of the sidebands increases with the cavity detuning, ultimately limiting the range of parameters over which the CSs can exist. In addition to improving our fundamental understanding of CS behaviour in longitudinally non-uniform resonators, our findings could impact on the design of future systems where dispersion management is leveraged to realise ultrashort CSs.

\section{Experimental setup and numerical model}

\begin{figure}[htb]	
\centering
		\includegraphics[width=0.7\linewidth]{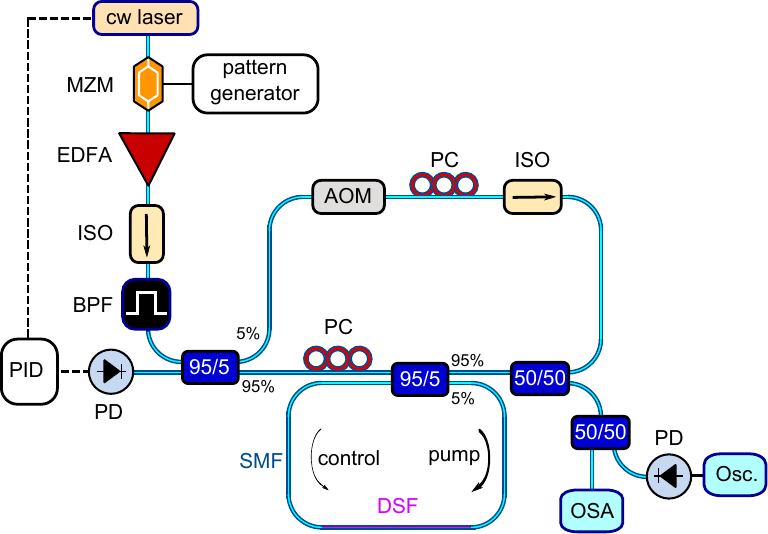}
		\caption{Schematic illustration of our experimental setup. MZM, Mach-Zehnder amplitude modulator; EDFA, Erbium-doped fiber amplifier; BPF, band-pass filter; PC, polarization controller; PD, photodetector; AOM, acousto-optic modulator; SMF, single-mode fiber; DSF, dispersion-shifted fiber; PID, proportional-integral-derivative controller; Osc., oscilloscope; OSA, optical spectrum analyzer; ISO, isolator.}
		\label{Fig1}
\end{figure}

Our experiment [c.f. Fig.~\ref{Fig1}] is centred around a dispersion-managed fiber ring resonator consisting of a 90-m-long segment of standard single mode optical fiber (SMF) and a 5-m-long segment of dispersion shifted fiber (DSF). At 1550 nm, the fibers' GVD coefficients are ${\beta_{2,\mathrm{SMF}}\approx -21.1~\mathrm{ps^2/km}}$ and ${\beta_{2,\mathrm{DSF}}\approx 2.0~\mathrm{ps^2/km}}$, respectively, yielding a net cavity GVD of $\langle\beta_2\rangle\approx-19.9~\mathrm{ps^2/km}$. (Note that, in contrast to refs.~\cite{jang_observation_2014, wang_universal_2017}, we have chosen to operate in the regime of comparatively weak dispersion management so as to minimize the impact of higher-order dispersion terms.) Although the fibers also have slightly different Kerr nonlinearity coefficients (the simulations below use $\gamma_\mathrm{SMF} = 1.2~\mathrm{W^{-1}km^{-1}}$ and $\gamma_\mathrm{DSF} = 1.8~\mathrm{W^{-1}km^{-1}}$), the resulting nonlinearity modulation plays a negligible role compared to the modulation of the GVD. We drive the resonator with flat-top nanosecond pulses carved from a narrow linewidth continuous wave (cw) laser with 1550~nm center wavelength. In addition to preventing the build-up of stimulated Brillouin scattering, the use of pulsed driving allows us to study CS behaviours over a much wider range of parameters than cw driving~\cite{anderson_observations_2016, anderson_coexistence_2017}. The driving pulses are carefully synchronized with the cavity round trip time, and launched into the resonator through a 95/5 coupler (input power coupling coefficient $\theta = 0.05$). The cavity has a measured finesse of about 50, corresponding to a fractional power loss $\rho\approx0.123$ per round trip, with losses arising predominantly from the coupling and the splices between the SMF and DSF segments. We monitor the CSs through the fourth port of the input coupler by means of a fast photodetector and an optical spectrum analyzer.

To sustain persisting CSs, the phase detuning $\delta_0$ between the driving laser and a cavity resonance must be actively stabilized~\cite{leo_temporal_2010}. We achieve this by using a method proposed in 1998 by Coen et al.~\cite{coen_experimental_1998}, where a small fraction of the pump is frequency-shifted using an acousto-optic modulator (AOM), and launched into the resonator in a direction opposite to the main pump. A proportional-integral-derivative (PID) controller is then used to actuate the driving laser frequency so as to hold the transmitted power of the backward propagating beam at a fixed level. A key advantage of this method is that, simply by adjusting the frequency of the electric signal applied on the AOM, the detuning experienced by the main pump can be locked at any value over an entire free-spectral range.

To corroborate and expand our experimental findings, we also numerically simulate the cavity dynamics using the infinite-dimensional Ikeda map:
\begin{align}
&E^{(m+1)}(0,\tau) = \sqrt{\theta}\,E_\mathrm{in}+\sqrt{1-\rho}\,E^{(m)}(L,\tau)e^{-i\delta_0}  \label{map_boundary} \\
&\frac{\partial E(z,\tau)}{\partial z} = i \sum_{k\geq 2}\frac{\beta_k(z)}{k!}\left(i\frac{\partial}{\partial \tau }\right)^k E  +  i\gamma(z) |E|^2 E  \label{map_GNLSE} \,.
\end{align}
Here, Eq.~\eqref{map_boundary} describes the coherent injection of the driving field into the resonator, whilst Eq.~\eqref{map_GNLSE} describes the propagation of the slowly varying electric field envelope $E^{(m)}(z,\tau)$ over a single pass through the fiber loop ($m$ is the round trip index). $E_\mathrm{in} = \sqrt{P_\mathrm{in}}$ describes the driving field with power $P_\mathrm{in}$ (our simulations assume cw driving), and $L$ is the total length of the cavity. Finally, $\beta_k(z)$ and $\gamma(z)$ are the usual dispersion and Kerr nonlinearity coefficients, respectively, with the $z$-dependence accounting for the presence of two different fibers. We include higher-order dispersion terms ($\beta_k$ with $k>2$) in Eq.~\eqref{map_GNLSE} for the sake of generality, but have found that in practice most of our results are well-described by the second-order dispersion alone. In fact, unless otherwise specified, the simulations shown below ignore higher-order dispersion terms (i.e., $\beta_k = 0$ for $k>2$). Note that the cavity dynamics can also be modelled using a piece-wise mean-field approach employed e.g. in~\cite{copie_competing_2016}, but we have opted to use the more general Ikeda map so as to fully capture the cavity periodicity.

\section{Results}

\begin{figure*}[t]	
		\includegraphics[width=0.8\linewidth]{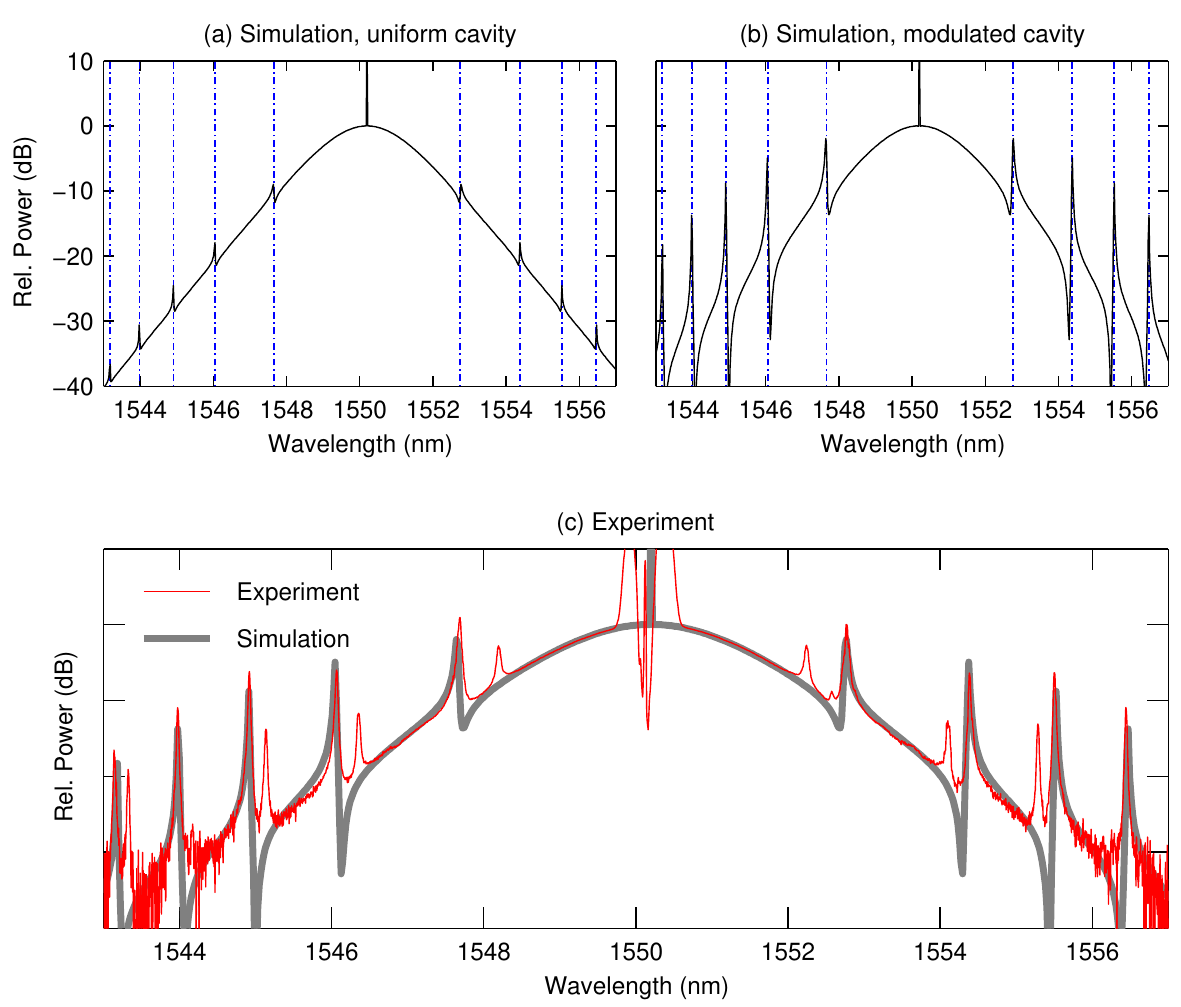}
		\caption{(a, b) Comparison of numerically simulated CS spectra in a longitudinally (a) uniform and (b) modulated resonator. The driving power $P_\mathrm{in} = 5~\mathrm{W}$ and detuning ${\delta_0 \approx 2.5~\mathrm{rad}}$. Higher-order dispersion terms are ignored ($\beta_k = 0$ for $k>2$). Vertical dash-dotted lines highlight the positions of the Kelly-like sidebands predicted by Eq.~\eqref{resonance}. (c) Red curve shows an experimentally measured spectrum of a CS in our dispersion-managed resonator. The corresponding simulation [from panel (b)] is also shown as a the gray curve.}
		\label{fig2}
\end{figure*}

We first show in Fig.~\ref{fig2}(a) a numerically simulated spectrum of a CS in a \emph{uniform} resonator, whose  GVD and nonlinearity coefficients correspond to the average values of our dispersion-managed cavity: ${\beta_2(z) = \langle \beta_2 \rangle\approx -19.9~\mathrm{ps^2/km}}$ and ${\gamma(z) = \langle \gamma \rangle  = 1.23~\mathrm{W^{-1}km^{-1}}}$ [other parameters as in our experiment described above; see also caption of Fig.~\ref{fig2}]. Although the simulated cavity is longitudinally uniform, weak Kelly-like sidebands appear atop the $\text{sech}^2$-shaped CS spectrum due to periodic losses caused by the 95/5 coupler. These sidebands appear at frequencies $\omega$ that are resonant with the driving field, and their positions can hence be predicted using the well-known phase-matching condition~\cite{luo_resonant_2015}:
\begin{equation}
\sum_{k\geq 2}\frac{\langle\beta_k\rangle}{k!}(\omega-\omega_0)^kL  = 2 \pi  n + \delta_0,
\label{resonance}
\end{equation}
where $n$ is an integer. As highlighted by the dash-dotted vertical lines in Fig.~\ref{fig2}(a), predictions from this phase-matching condition are in excellent agreement with the simulated sideband positions.

Next, in Fig.~\ref{fig2}(b), we show results from numerical simulations that explicitly take into account the presence of two different fibers [other parameters as in Fig.~\ref{fig2}(a)]. We see that, while the $\text{sech}^2$-shaped spectral envelope of the soliton is largely unchanged, the intensities of the resonant sidebands have increased by about 10~dB. Such strong sidebands are also clearly present in the corresponding \emph{experimentally} measured spectrum, shown as the solid red curve in Fig.~\ref{fig2}(c). (Note that, to improve signal-to-noise ratio, a band-stop filter was used to remove the central part of the spectrum.) These results show that dispersion management (i) can magnify the intensities of the resonant radiation sidebands but (ii) does not significantly affect their positions. We also highlight that the experimentally measured spectrum is in excellent agreement with the corresponding numerical simulation [gray curve in Fig.~\ref{fig2}(c); same data as in Fig.~\ref{fig2}(b)].

Although the experimentally measured spectrum in Fig.~\ref{fig2}(c) shows very good overall agreement with the corresponding numerical simulation, there is one notable discrepancy. Specifically, the experiment shows small auxiliary sidebands that are not present in our simulations. Additional measurements (not shown) reveal these sidebands to be orthogonally polarised with respect to the rest of the spectrum, allowing us to conclude that they arise from the nonlinear coupling of the two principal polarization modes of our cavity, as recently reported in~\cite{wang_universal_2017}. Because of the scalar nature of Eqs.~\eqref{map_boundary} and~\eqref{map_GNLSE}, these sidebands are naturally not present in our simulations.

\begin{figure}[b]	
		\includegraphics[width=0.8\linewidth]{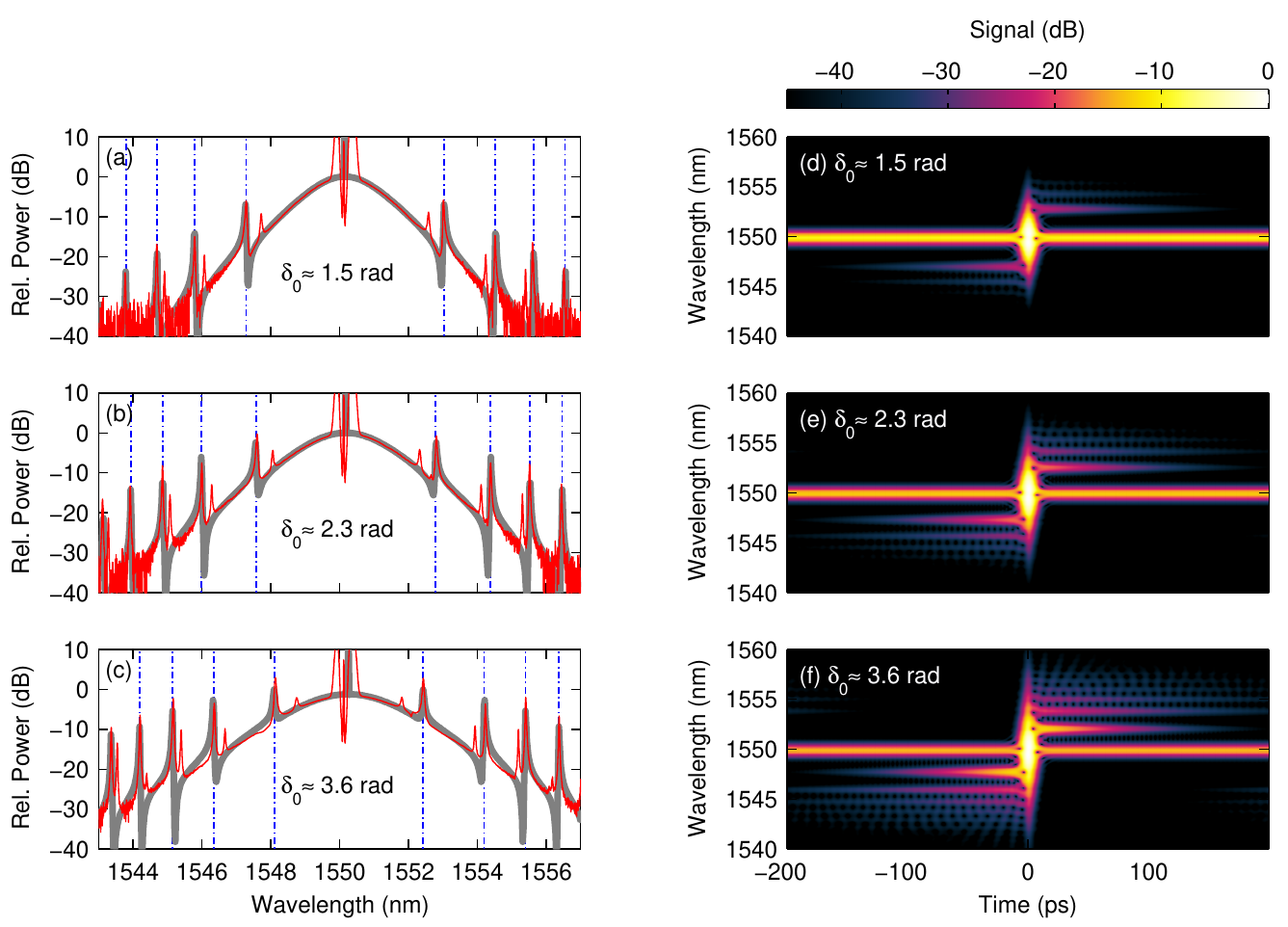}
		\caption{(a--c) CS spectra at three different detunings: (a) ${\delta_0 \approx 1.5~\mathrm{rad}}$, (b) $\delta_0 \approx 2.3~\mathrm{rad}$, and (c) $\delta_0 \approx 3.6~\mathrm{rad}$. Red and gray curves show results from experimental measurements and corresponding numerical simulations, respectively. For each detuning, the driving power $P_\mathrm{in}\approx 5~\mathrm{W}$. (d--f) Spectrograms of the intracavity fields corresponding to the numerically simulated spectra in (a--c).}
		\label{fig3}
\end{figure}

Equation~\eqref{resonance} predicts that the sidebands will shift closer to the pump frequency as the detuning $\delta_0$ increases~\cite{luo_resonant_2015}. Compounded by the fact that the CS duration decreases with detuning~\cite{coen_universal_2013}, we may expect the sideband amplitudes to increase, as more spectral energy is available at the resonant frequencies. To quantitatively elaborate on these notions, we first recall that the 3 dB spectral width of a hyperbolic secant-shaped Kerr CS~\cite{coen_universal_2013} can be approximated as
\begin{equation}
\Omega_\mathrm{CS}\approx \sqrt{\frac{2\delta_0}{|\langle\beta_2\rangle|L}}.
\end{equation}
For CSs to exist, $\delta_0 > 0$ and $\langle\beta_2\rangle < 0$, and so the frequency shift $\Omega_\mathrm{pm} = \omega_\mathrm{pm}-\omega_0$ of the first-order Kelly-like sidebands (given by Eq.~\eqref{resonance} with $n = -1$) is approximately
\begin{equation}
\Omega_\mathrm{pm} = \pm\sqrt{\frac{4\pi-2\delta_0}{|\langle\beta_2\rangle|L}}.
\end{equation}
Combining the two expressions yields:
\begin{equation}
\frac{\Omega_\mathrm{pm}}{\Omega_\mathrm{CS}} = \pm\sqrt{\frac{2\pi}{\delta_0} - 1}.
\label{pmcond}
\end{equation}
From this expression, we indeed see clearly that (i) the resonant sidebands shift closer to the pump as the detuning increases ($\Omega_\mathrm{pm}\rightarrow 0$ as $\delta_0\rightarrow 2\pi$), and (ii) the spectral amplitude at the resonant frequency increases with detuning [$\tilde{E}(\Omega_\mathrm{pm})\propto \text{sech}(\Omega_\mathrm{pm}/\Omega_\mathrm{CS})$]. It is worth highlighting that Eq.~\eqref{pmcond} does not depend on the net GVD: the CS bandwidth and the frequency-shift of the resonant sideband both scale as $|\langle\beta_2\rangle|^{-1/2}$.

Our experiments and simulations confirm the detuning-dependence predicted above, as shown in Figs.~\ref{fig3}(a--c). Here we display measured and simulated CS spectra at three different detunings (as indicated), together with theoretically predicted phase-matching frequencies. We see clearly how the sidebands --- whose positions are very well predicted by Eq.~\eqref{resonance} --- shift towards the pump and grow in amplitude as the detuning increases. It is interesting to note that, in the time domain, the sidebands correspond to extended temporal features that are leading (blue-shifted sidebands) or trailing (red-shifted sidebands) the CS due to (anomalous) dispersion. This is illustrated in Figs.~\ref{fig3}(d--f), where we show the \emph{spectrograms} --- i.e., the time-frequency representations --- of the numerically simulated fields whose spectra are shown in Figs.~\ref{fig3}(a--c). The interference between the frequency-shifted sidebands and the cw-background on top of which the CSs sit gives rise to strong oscillatory tails that emanate from the CS. As expected~\cite{wang_universal_2017}, these tails can give rise to very robust bound soliton states, which we have routinely observed in our experiments.

\begin{figure}[b]	
		\includegraphics[width=0.8\linewidth]{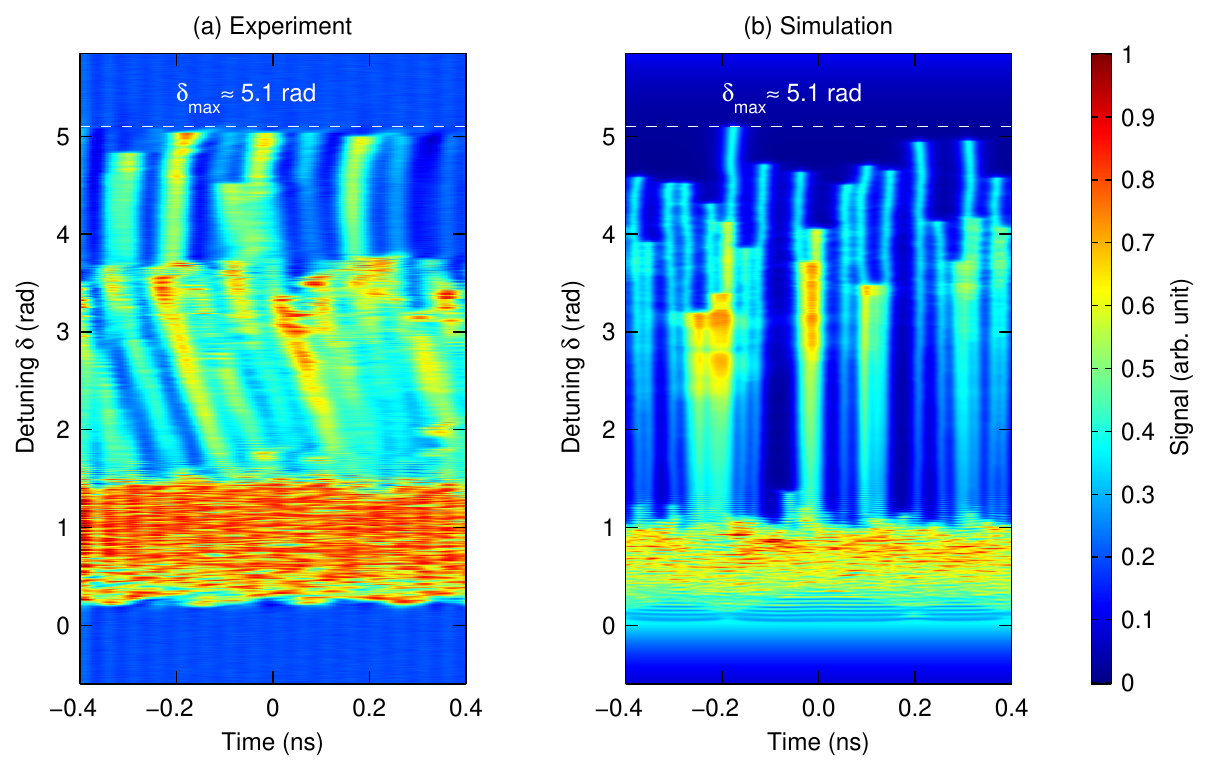}
		\caption{(a) Measured and (b) simulated intracavity dynamics as the detuning is scanned over one cavity free-spectral range in about 2000 round trips. The simulations use experimental parameters, with driving power $P_\mathrm{in} = 5.5~\mathrm{W}$. To facilitate visualisation, the simulated data has been convolved with a 10~ps response function. The dashed horizontal line indicates the limit detuning $\delta_\mathrm{max}\approx5.1~\mathrm{rad}$ at which the last CSs cease to exist.}
		\label{fig4}
\end{figure}

The strongest sidebands in Fig.~\ref{fig3}(c) exceed the peak spectral intensity of the CS by several dB. We may intuitively expect that, if the sidebands grow sufficiently large, they may perturb the soliton to the extent that it can no longer exist. Because the sidebands increase with increasing cavity detuning, this suggests that the emission of resonant radiation may diminish the range of detunings over which the CSs can exist. We have tested this hypothesis by performing experiments similar to those in~\cite{wang_stimulated_2018}. Specifically, we slowly scan the driving laser frequency across a single cavity resonance, and continuously record the intensity coupled out of the resonator on a real-time oscilloscope. The out-coupled field passes through a band-pass filter off-set from the 1550~nm driving wavelength prior to detection so as to improve the signal-to-noise ratio and to better isolate the solitons' creation and annihilation dynamics~\cite{luo_spontaneous_2015,anderson_observations_2016}. From the acquired data, we can extract the maximum detuning $\delta_\mathrm{max}$ beyond which CSs no longer exist. These measurements are repeated for a range of driving power levels $P_\mathrm{in}$, and compared with the theoretical prediction derived for a homogeneous, high-finesse resonator~\cite{herr_temporal_2014, barashenkov_existence_1996}:
\begin{equation}
\delta_\mathrm{max,\,t} \approx \frac{\pi^2}{8}\frac{\langle\gamma\rangle P_\mathrm{in}\theta L}{\alpha^2},
\label{deltalimit}
\end{equation}
where $\alpha$ describes the resonator loss. We must note that Eq.~\eqref{deltalimit} is derived in the mean-field approximation of the full cavity map described by Eqs.~\eqref{map_boundary} and~\eqref{map_GNLSE}. In this approximation, there are two well-justified yet slightly different expressions for the loss coefficient $\alpha$: $\alpha_1 = \rho/2$ and $\alpha_2 = \pi/\mathcal{F}$. As expected, in the limit $\mathcal{F}\rightarrow\infty$, we find $\alpha_1 = \alpha_2$; outside this limit, extensive simulations show that best agreement between the full cavity map and the mean-field prediction is obtained when choosing $\alpha = (\alpha_1 + \alpha_2)/2$. Accordingly, below we compare our full cavity map experiments and simulations with Eq.~\eqref{deltalimit} using $\alpha = 0.063$.

\begin{figure}[b]	
		\includegraphics[width=0.8\linewidth]{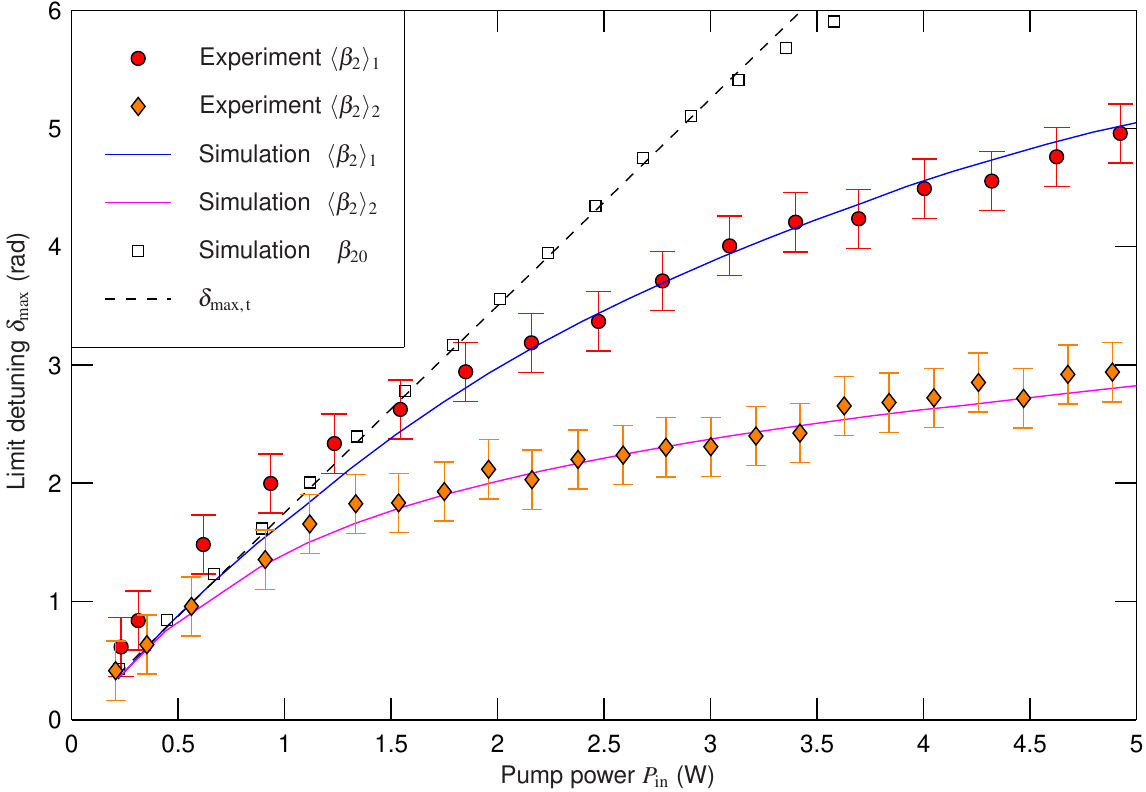}
		\caption{Comparison of limit detunings at which CSs cease to exist. Red circles and orange diamonds correspond to experiments performed in two different dispersion-managed resonators with average GVD $\langle\beta_2\rangle_1 \approx -19.9~\mathrm{ps^2/km}$ and $\langle\beta_2\rangle_2 \approx -13.7~\mathrm{ps^2/km}$, respectively. The solid blue and magenta curves show corresponding simulation results. Dashed line is the theoretically predicted limit detuning given by Eq.~\eqref{deltalimit}, while the open squares were obtained from simulations of a uniform resonator with GVD $\beta_2(z) = \beta_{20} = \langle\beta_2\rangle_1$.}
		\label{fig5}
\end{figure}

Figure~\ref{fig4}(a) and (b) respectively show examples of experimentally measured and numerically simulated evolutions of the intracavity intensity profile as the detuning is scanned over one cavity free-spectral range (during about 2000 round trips). We can see how CSs emerge from an extended modulation instability pattern at a detuning $\delta_0\approx 1.5~\mathrm{rad}$~\cite{herr_temporal_2014,luo_spontaneous_2015}. The solitons exhibit curved trajectories in the experimentally measured trace due to stimulated Raman scattering (SRS)~\cite{anderson_coexistence_2017, wang_stimulated_2018}; this effect does not manifest itself in our simulations as our theoretical model ignores SRS. A large number (but not all) of the CSs are seen to disappear at detunings between  $\delta_0\approx 3-4~\mathrm{rad}$. We believe this is because the strong interactions mediated by the Kelly-like sidebands render (some) bound soliton states unstable, allowing only widely-spaced CSs to persist. The last CSs are finally seen to disappear at $\delta_\mathrm{max}\approx 5.1~\mathrm{rad}$ --- a value significantly smaller than $\delta_\mathrm{max,t}\approx 9.6~\mathrm{rad}$ predicted for a uniform cavity by Eq.~\eqref{deltalimit}.

To enable a more comprehensive comparison, we have performed experiments and simulations over a wide range of pump powers. Moreover, in addition to the resonator described above, we have also considered (both experimentally and numerically) another resonator made out of a 65-m-long-segment of SMF and 30-m-long segment of DSF (average GVD $\langle\beta_2\rangle\approx-13.7~\mathrm{ps^2/km}$; other parameters, including finesse, similar as before). Figure~\ref{fig5} summarizes our results. Here we show the limit detunings extracted from our experiments (red circles and orange diamonds) and simulations (blue and magenta solid curves), together with the theoretical limit detunings predicted by Eq.~\eqref{deltalimit} (black dashed curve). For completeness, we also show the limit detunings extracted from numerical simulation of a uniform cavity whose parameters are equal to the averaged parameters of one of our dispersion-managed resonators ($\beta_2(z) = -19.9~\mathrm{ps^2/km}$, black squares).

Our experiments and simulations show that the range of soliton existence is diminished in dispersion-managed resonators. Moreover, it is evident that, out of the two dispersion-managed resonators considered, the range of CS existence is more severely limited in the one with the smaller average GVD, i.e., the one exhibiting ``stronger'' dispersion management. We must emphasize that these findings cannot be explained by SRS, which has also recently been shown to limit CS existence at large detunings~\cite{wang_stimulated_2018}. Indeed, additional simulations (not shown here) reveal that (i) inclusion of SRS does not materially change the results in Fig.~\ref{fig5} and that (ii) SRS alone (in the absence of dispersion management) significantly overestimates the range of CS existence. These observations do not contradict ref.~\cite{wang_stimulated_2018}: our parameters are simply such that SRS is overshadowed by the effect of dispersion management.

\begin{figure}[b]	
		\includegraphics[width=0.8\linewidth]{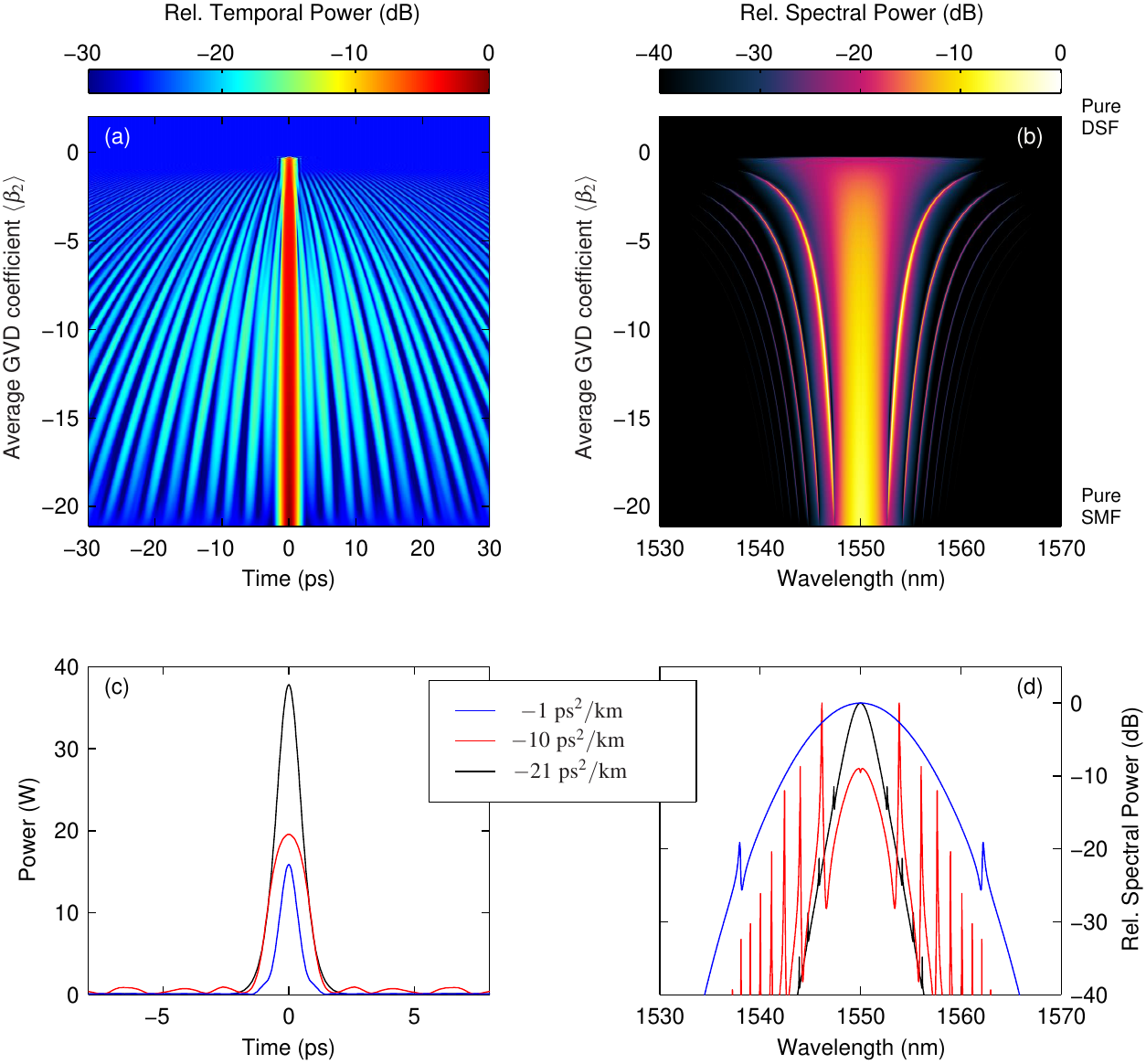}
		\caption{Numerical simulation results, showing the (a) temporal and (b) spectral profile of a single CS as a function of the average cavity GVD $\langle\beta_2\rangle$. Other cavity parameters, including the total cavity length, kept constant: $L_\mathrm{tot} = 95~\mathrm{m}$; $P_\mathrm{in} = 5~\mathrm{W}$; $\delta_0 = 2~\mathrm{rad}$; $\theta = 0.05$; $\rho = 0.123$. (c, d) Simulated CS (c) intensity profiles and (d) spectra for selected values of $\langle\beta_2\rangle$ as indicated.}
		\label{fig6}
\end{figure}

The experimental and numerical data presented in Fig.~\ref{fig5} suggest that the CSs are more strongly perturbed in the resonator with lower average GVD. This is of course not particularly surprising given the increased amount of normally dispersive DSF. To gain more insights, we have calculated the CS spectral and temporal characteristics as a function of average GVD $\langle\beta_2\rangle$. This was achieved by adjusting the lengths of the SMF and DSF segments in our simulations while keeping the total cavity length (and all other parameters) constant. Results are summarized in Fig.~\ref{fig6}. Here the false colour plots show the temporal [Fig.~\ref{fig6}(a)] and spectral [Fig.~\ref{fig6}(b)] intensity profiles of the CS in logarithmic scale as a function of the net cavity GVD, while the line plots in Fig.~\ref{fig6}(c) and (d) show selected profiles in more detail. Note that the bottom-most and top-most profiles in the false colour plots correspond to the scenarios where the entire cavity is made out of SMF and DSF, respectively. Note also that the simulated profiles shown are extracted immediately after the DSF segment. Because the pulses temporally broaden in the normally dispersive DSF, the intensity profiles shown do not capture the pulse shortening that can be achieved in a dispersion-managed system~\cite{bao_stretched_2015}. Considering the case with $\langle\beta_2\rangle = -1~\mathrm{ps^2/km}$ [blue curves in Fig.~\ref{fig6}(c) and (d)] as an illustrative example, our simulations show that the minimum duration achieved (about half-way through the SMF segment) is about 380~fs, in stark contrast with the 880-fs-duration at the end of the DSF segment.

There are several conclusions to be drawn from Fig.~\ref{fig6}. First, as expected, CSs only exist when the net GVD is negative, i.e., $\langle\beta_2\rangle < 0$. Secondly, in accordance with our experimental observations, we see how the introduction of dispersion management magnifies the Kelly-like resonant radiation sidebands and corresponding time-domain modulations. Interestingly, for sufficiently large lengths of DSF (small average GVD), the resonant radiation sidebands effectively disappear. (Similar behaviour was numerically reported also in~\cite{bao_stretched_2015}.) We speculate this is because the overall spectrum transforms from the $\text{sech}^2$ profile characteristic to a Kerr CS towards a Gaussian shape (parabolic on logarithmic scale), such that the scaling behind Eq.~\eqref{pmcond} no longer holds. In this regime, a decreasing GVD shifts the phase-matched sidebands away from the pump more rapidly than the bandwidth of the CS expands, resulting in less energy available to seed the sidebands.

\begin{figure}[tb]	
		\includegraphics[width=0.8\linewidth]{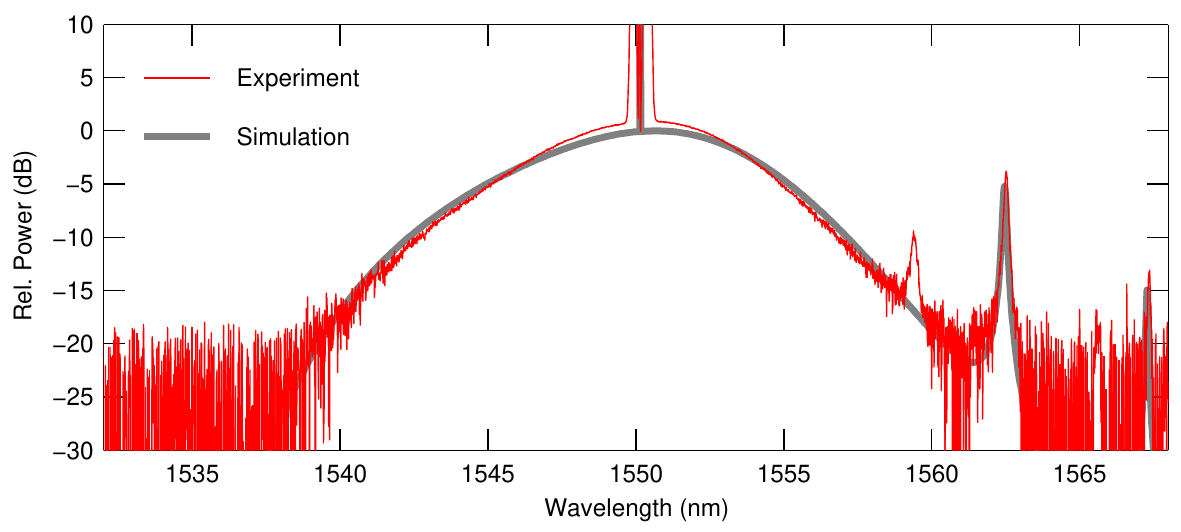}
		\caption{Experimentally measured (red) and numerically simulated (gray) CS spectra in a dispersion-managed resonator with $\langle\beta_2\rangle\approx -0.8~\mathrm{ps^2/km}$. Driving power $P_\mathrm{in} = 5~\mathrm{W}$ and detuning $\delta_0 = 1.3~\mathrm{rad}$. The simulations assume the DSF and SMF to have third-order dispersion coefficients $\beta_{3,\mathrm{SMF}} = 0.1~\mathrm{ps^3/km}$ and $\beta_{3,\mathrm{SMF}} = 0.16~\mathrm{ps^3/km}$, respectively.}
 		\label{fig7}
\end{figure}

To experimentally probe this latter simulation result, we built a dispersion-managed cavity with a very low average GVD $\langle\beta_2\rangle\approx -0.8~\mathrm{ps^2/km}$ ($L_\mathrm{SMF} \approx 11~\mathrm{m} $, $L_\mathrm{DSF} \approx 79~\mathrm{m}$; other parameters, including finesse, similar as before). Figure~\ref{fig7} shows an example of the spectrum measured for a CS sustained in this resonator. Whilst we see that the spectrum does indeed exhibit a more Gaussian-like profile compared to e.g. those shown in Fig.~\ref{fig3}, we also still see comparatively strong resonant radiation sidebands. These sidebands arise because of third-order dispersion (which was neglected in results shown in Fig.~\ref{fig6}), as confirmed by numerical simulations (gray curve in Fig.~\ref{fig7}). This is not unexpected: as the net GVD approaches zero, higher-order dispersion terms become more pronounced and can no longer be neglected. Thus, although strong dispersion management can to some extent mitigate the generation of resonant radiation sidebands, complete suppression would require additional management of higher-order dispersion terms.

\section{Discussion and Conclusion}

Before closing, we briefly discuss the relevance of our findings to high-Q microresonator systems. As proposed in~\cite{bao_stretched_2015}, dispersion management could facilitate the generation of broadband optical frequency combs in such systems, and experimental realizations of high-Q devices with longitudinal parameter variations have already been reported~\cite{kordts_higher_2016, huang_quasi-phase-matched_2017}. In uniform microresonators, CSs do not efficiently excite Kelly-like resonant radiation sidebands for two reasons. First, because of the devices' low losses, the solitons experience negligible evolution over one round-trip, and hence shed little linear radiation that could build up to strong sidebands. Secondly, in typical microresonator experiments $\delta_0\ll 1~\mathrm{rad}$, such that the resonant sidebands are far away from the pump relative to the 3-dB CS bandwidth [see Eq.~\eqref{pmcond}]. Efficient sideband generation is therefore not to be expected, as little spectral energy is available at the phase-matched frequency. Because dispersion management does not affect the sideband frequencies (compared to a uniform resonator with the same average GVD), we would not expect the solitons to be significantly perturbed under typical microresonator conditions. Of course, the situation changes entirely if $\delta_0$ is of the order of unity (as in our fiber resonator experiments), in which case we would expect dispersion management to stimulate the growth of strong sidebands.

In conclusion, we have experimentally and numerically studied the effects of dispersion management on temporal Kerr CSs. We have shown that, in a manner analogous to the amplification of parametric instabilities in resonators with net normal GVD~\cite{copie_competing_2016, conforti_parametric_2016, copie_dynamics_2017}, dispersion management can magnify the resonant radiation sidebands emitted by CSs. Our experiments and simulations also show that the magnification of the radiation sideband can limit the range of detunings over which the solitons exist. All our experimental observations are in good agreement with corresponding numerical simulations. In addition to elucidating the behaviour of Kerr CSs in longitudinally modulated resonators, our work improves our general understanding of CSs in the presence of higher-order perturbations.

\begin{acknowledgments}
The authors wish to acknowledge the Centre for eResearch at the University of Auckland for their help in facilitating this research. We also acknowledge financial support from the Marsden Fund and the Rutherford Discovery and James Cook Research Fellowships of the Royal Society of New Zealand.
\end{acknowledgments}

\end{document}